\magnification1200

%\rightline{KCL-MTH-14-05}
%\rightline{hep-th/yymmnnn}

\vskip 2cm
\centerline
{\bf  The non-linear  dual gravity equation of motion in eleven dimensions}
\vskip 1cm
\centerline{Keith Glennon and Peter West}
\centerline{Department of Mathematics}
\centerline{King's College, London WC2R 2LS, UK}
\vskip 2cm
%\centerline{and}
%\vskip 0.5cm
%\centerline{Peter West,}
%\centerline {,}
%\centerline{}

\rightline {$E_{11}$ knows best.}
\vskip1cm
\leftline{\sl Abstract}
We derive the non-linear dual graviton equation of motion in eleven dimensions in the context of E theory.  
\vskip2cm
\noindent

\vskip .5cm

\vfill
\eject

\medskip
{{\bf 1. Introduction}}
\medskip

The classic paper of Montonen and Olive [1] suggested that spontaneously broken SU(2) Yang-Mills theory  with a triplet Higgs  possessed a duality symmetry that transformed  the electric particles into the magnetic particles.  Duality symmetries also play an important role in supergravity theories 
but in these theories one finds that the duality transformations often exchange form fields of different rank instead of the vector potentials in the case 
considered by Montonen and Olive. The non-linear realisation of 
$E_{11} \otimes_s l_1$ is a theory that contains all  the maximal supergravity theories, depending on the decomposition of $E_{11}$ that one takes [2,3]. Contained in this symmetry  are the $E_{11-D} $ (Cremmer-Julia) symmetry [4] of the maximally supergravity theories in $D$ dimensions as well as the Sl(2,R) symmetry [5] of IIB supergravity. However, the $E_{11}$ symmetry also contains symmetries that transform  the fields of different spin into each other [6]. For example,  in eleven dimensions the symmetry  transforms the graviton field into the three form which also transforms into  the six form field etc. The non-linear realisation of $E_{11} \otimes_s l_1$ contains an infinite number of fields quite a few of these correspond to the infinitely many  different ways of describing the some on-shell state  [7,8]. For example the on-shell states described by the three form can also be described by a six form field but also by the field $A_{a_1\ldots a_9, b_1b_2b_3}$ and indeed any such  field that has any number of antisymmetric blocks of nine indices. The $E_{11}$ symmetry contains not only the known duality transformations but also an infinite number of new duality symmetries that transform these duality equivalent field descriptions  into each other. 
\par
The non-linear realisation of $E_{11} \otimes_s l_1$ contains the usual field of gravity which satisfies Einstein equation in the presence of the three form field. This statement holds if one keeps  only the lowest level  coordinates of spacetime which are just those of our familiar spacetime. The non-linear realisation of $E_{11} \otimes_s l_1$ also contains the field of dual gravity. A field that described dual gravity was first proposed by Curtright in five dimensions, it was of the form 
$h_{a_1a_2 ,b}$ [9] and it was later proposed that  the  field $h_{a_1\ldots a_{D-3}, b}$ described dual gravity  in $D$ dimensions [10].  In reference [6] the equation of motion of this field was found  and it was shown to  describe the degrees of freedom of gravity in $D$ dimensions {\bf at the linearised level}. 
\par
To find the {\bf non-linear } equation obeyed by the dual graviton  has proved much more difficult. Indeed a no go theorem has been proved [11]. However, discussions of duality in the context of gravity require such an equation to exist. Since the non-linear realisation of $E_{11} \otimes_s l_1$ contains the field of dual gravity one should be able to deduce its equation of motion  from the non-linear realisation.  In particular in eleven dimensions one finds that the $E_{11}$ variation of the six form equation of motion  contains the dual graviton equation  of motion. The correct linearised dual gravity equation of motion was derived in this way at the linearised level in reference [16] while the dual graviton equation at the non-linear level  was derived in reference [12]. This  latter derivation did not appear to lead to a unique dual gravity equation of motion  but one that was ambiguous up to certain types of terms which were albeit it of a very restricted type. In reference [12] this ambiguity was  apparently resolved by insisting on the additional requirement of diffeomorphism invariance. Such additional requirements were not  used in any  other $E_{11}$ papers when deriving the equations of motion This paper also found the duality equation that is first order in spacetime derivatives which  relates the gravity field to the dual gravity field. 
\par
It was apparent from reference [12] how  the no go theorem of reference [11] is circumvented. Although the field $h_{a_1\ldots a_8, b}$ on its own does correctly describe  gravity at the linearised level this is not the case at the non-linear level. However, the dual gravity equation  of motion that follows from the non-linear realisation of $E_{11} \otimes_s l_1$ contains both the usual gravity field as well as the dual gravity field. This is to be expected. It is well known that the non-linear equation of motion of the six form must also contains the three form while the duality related equation of motion of the three form just contains the three form field. 
\par
 Recently  the non-linear realisation the semi-direct product of $A_{1}^{+++}$ and it first fundamental (vector) representation $l_1$, denoted as $A_{1}^{+++}\otimes_s l_1$, was constructed [13]. This theory contains the graviton $h_{ab}$, at level zero, and the dual graviton $\tilde h_{ab}=\tilde h_{(ab)}$ at the next level as well as higher level fields. The gravity and dual gravity equations of motion as well as the duality relation that relates the two fields were found.  The dual graviton  equation of motion was  essentially unique provided we demanded that it had  the same index structure as the dual graviton field, that is, it was symmetric in its two indices. Of course this is not so much a demand as a necessity. The results of this paper  made it clear that while many of the results in reference [12] were correct the equation of motion of the dual gravity was not correct. 
 \par
 In this paper we revisit the calculations of reference [12] and consider again the $E_{11}$ variation of the six form equation of motion. We demand that the dual graviton equation $E_{a_1\ldots a_8, b}$ carries the same index structure as the dual graviton field $h_{a_1\ldots a_8, b}$, that is, it obeys the condition $E_{[a_1\ldots a_8, b]}=0$. The dual graviton equation is then essentially unique without the need for any extraneous additional requirements.

 %%%%%%%%%%%%%%%%%%%%%%%%%%%%%%%%%%%%%%%%%%%%%%%%%%%%%%%%%%%%%%%%%%%%%%%
 \medskip
{{\bf 2. The calculation }}
\medskip
The construction of the non-linear realisation of  $E_{11} \otimes_s l_1$ in eleven dimensions has been much studied and so we will not repeat it here. The reader is referred to references [2,3]  and [12], the review of reference [14]  and the book of reference [15] for the details. We will now just recall our  essential starting points from reference [12]. The Cartan forms, up to level three, are given by [2,3]
$$
G_c {}_{, a}{}^b = (\det e)^{{1\over 2}}   e_c{}^\tau e_a{}^\rho \partial_\tau  e_\rho{}^b,
$$ 
$$
{ G}_{c,a_1 a_2 a_3} = (\det e)^{{1\over 2}}e_c{}^\tau 
e_{[a_1}{}^{\mu_1} e_{a_2}{}^{\mu_2}  e_{a_3]}{}^{\mu_3} \partial_{\tau} A_{\mu_1\mu_2\mu_3}, 
$$
$$
{ G}_{c , a_1\ldots a_6 }= (\det e)^{{1\over 2}} e_c{}^\tau  e_{[a_1}{}^{\mu_1}\ldots e_{a_6]}{}^{\mu_6}(\partial_{\tau} A_{\mu_1\ldots \mu_6}-A_{\mu_1\mu_2\mu_3} \partial_{\tau}A_{\mu_4\mu_5 \mu_6})
$$
$$
G_{c}{}_{, a_1\ldots a_8,b} =   (\det e)^{{1\over 2}}  e_c{}^\tau e_{[a_1}{}^{\mu_1}\ldots e_{a_8]}{}^{\mu_8}e_{b}{}^{\nu}   (\partial_\tau h_{\mu_1\ldots \mu_8,\nu}
-A_{\mu_1\mu_2 \mu_3}\partial_\tau A_{\mu_4\mu_5\mu_6} A_{\mu_7\mu_8 \nu}
$$
$$
+2 \partial_\tau A_{\mu_1\ldots \mu_6} A_{\mu_7\mu_8 \nu}
+2\partial_\tau A_{\mu_1\ldots \mu_5 \nu} A_{\mu_6\mu_7  \mu_8})
\eqno(2.1)$$
where the vierbein is given in terms of the field $h_a{}^b$  by $e_\mu{}^a \equiv (e^h)_\mu{}^a$
\par
They are inert under rigid $E_{11}\otimes_s l_1$ transformations but transform under $I_c(E_{11})$ transformations as [2,3]
$$
\delta G_{a}{}^{b}=18\, \Lambda^{c_1c_2 b }G_{c_1c_2 a}
-2\, \delta_a ^{b}  \Lambda^{c_1c_2 c_3}G_{c_1c_2 c_3},\ 
\eqno(2.2)$$
$$
\delta G_{a_1a_2a_3}=-{5!\over 2} G_{b_1b_2b_3 a_1a_2a_3}
\Lambda^{b_1b_2 b_3} -6\, G_{(c [a_1 |) } \Lambda_{c}{}_{|a_2a_3]}
\eqno(2.3)$$
$$
\delta G_{a_1\ldots a_6}=2 \Lambda_{[ a_1a_2a_3}G_{a_4a_5a_6 ]}
-112\, G_{b_1b_2b_3 [ a_1\ldots a_5,a_6]}\Lambda^{b_1b_2b_3}
+112\, G_{b_1b_2 a_1\ldots a_5a_6, b_3 }\Lambda^{b_1b_2b_3}
$$
$$
= 2 \Lambda_{[ a_1a_2a_3}G_{a_4a_5a_6 ]}
-336 \, G_{b_1b_2b_3 [ a_1\ldots a_5,a_6]}\Lambda^{b_1b_2b_3}\
\eqno(2.4)$$
$$
\delta G_{a_1\ldots a_8,b}=-3\, G_{[ a_1\ldots a_6}\Lambda_{a_7a_8] b}
+3 \,G_{[ a_1\ldots a_6}\Lambda_{a_7a_8 b]}
\eqno(2.5)$$
\par
The above formulae are true when the Cartan forms are written as forms, for example $G_a{}^b = dz^\Pi G_{\Pi ,}{}_a{}^b$. We will convert  their  first world volume index     into a tangent index by using the formula 
$G_A{}_{, \underline \alpha }=E_A{}^\Pi G_\Pi{}_{, \underline \alpha }$  where $E_\Pi{}^A$ is the vierbein on the spacetime encoded in the non-linear realisation. Under the $I_c(E_{11})$ transformations this first  index  on the Cartn forms transforms as 
$$
\delta G_{a, \bullet}= -3G^{b_1b_2}{}_{,\bullet}\ 
\Lambda_{b_1b_2 a},
\quad \delta G^{a_1a_2}{}_{, \bullet}= 6\Lambda^{a_1a_2
b}  G_{b,}{}_{\bullet}, \ \ldots 
\eqno(2.6)$$
Thus  the Cartan forms $G_A{}_{, \underline \alpha }$  transform under equation (2.6) on their first $(l_1)$ index and their $E_{11}$ indices transform as in equations (2.2) to (2.5). 
\par
In the papers in references [2,3,12] etc  we have found  the equations of motion only  to lowest order in derivatives of the spacetime coordinates, that is,  just with  derivatives with respect to the usual coordinates of spacetime. However, according to equation  (2.6) Cartan forms with level one derivatives can transform into Cartan forms with derivatives with respect to the level zero coordinates. Consequently  to find the equations of motion to lowest order in the spacetime derivatives we need the equations that we are varying to contain  the required derivatives with respect to the level one coordinates. As in the previous papers we denote the former quantities by just capital letters and the later by calligraphic letters. 
In reference [12] the six form equation of motion  was found by varying the three form equation of motion, its precise form is given by 
$$
\hat {\cal E}^{a_1\ldots a_6}= e_{\mu_1}{}^{ [ a_1} \ldots e_{\mu_6}{}^{a_6 ]} \bigg(\partial_{\nu} \big((\det e)^{{1\over 2}} G^{[\nu , \mu_1\ldots \mu_6]}\big) 
-8\, \partial_\nu ( (\det e )^{{1\over2}} G^{\tau_1 \tau_2 , \nu \mu_1\ldots \mu_6 }{}_{\tau_1, \tau_2})
$$
$$
+{1\over 7} (\det e )^{-{1\over2}}
 \partial^{\mu_1\mu_2}
( (\det e )^{{1\over2}} G^{\mu_3, \mu_4 \mu_5\mu_6}) 
-{72} \, (\det e )^{{1\over2}} G^{[\nu , \mu_1\ldots \mu_6 \sigma_1 \lambda]} {}_{, \tau } Q^{\tau}{}_{ \sigma_1 }{}_{, \nu \lambda}
$$
$$
-36 \, e_{\tau_1}{}^{b_1} e_{\tau_2} {}^{b_2} e_{\rho_1 b_1}e_{\rho_2 b_2 } \partial^{\rho_1\rho_2} ( (\det e )^{{1\over2}} G^{[ \nu , \mu_1\ldots \mu_6 \tau_1\tau_2 ]}{}_{ , \nu})\bigg)
$$
$$
-3 G^{c_1c_2}{}_{,c_1c_2}{}_{ e} G^{[e , a_1\ldots a_6 ]}
-18\, G^{c [a_1 |} {}_{, c d_1d_2} G^{[ d_1, d_2 | a_2 \dots a_6]]} =0.
\eqno(2.7)$$
We have put a hat on the symbol for the six form equation as this  will not be our final result.  
Its   $I_c(E_{11})$  variation can be written as [12]
$$
\delta \hat {\cal E}^{a_1\ldots a_6 }= 432\,\Lambda_{c_1c_2c_3}\,\hat E^{a_1...a_6c_1c_2,\,c_3}+{8\over 7}  \Lambda ^{[a_1a_2a_3} E^{a_4a_5a_6]}
$$
$$
+{2\over 105} G_{[e_5 ,  c_1c_2c_3]} \epsilon^{a_1\ldots a_6e_1\ldots e_5} E_{e_1\ldots e_4}\Lambda^{c_1c_2c_3}
$$
$$
-{3\over 35} G_{[ e_5, e_6 c_1c_2]}\Lambda^{c_1c_2[a_1}
\epsilon^{a_2\ldots a_6 ] e_1\ldots e_6} E_{e_1\ldots e_4}
+ {1\over 420}\,\epsilon_{c_1}{}^{a_1...a_6b_1...b_4}\,\omega_{c_2,\,b_1b_2}\,E_{c_3,\,b_3b_4}\,\Lambda^{c_1c_2c_3}
\eqno(2.8)$$
where $E_{a,bc}$ is the gravity-dual gravity duality relation derived, for example, in reference [12]  to be 
$$
E_{a,b_1 b_2} = (\det e)^{1 \over 2} \omega_{a,b_1b_2} - {1 \over 4} \varepsilon_{b_1b_2}{}^{c_1 \ldots c_9} G_{c_1,c_2 \ldots c_9,a} \dot{=} 0
\eqno(2.9)$$
where the spin connection is given by  
$$
(\det e)^{1 \over 2} \omega_{a,b_1 b_2} = - G^S_{b_1,b_2a} + G^S_{b_2,b_1a} + G_{a,[b_1 b_2]}, \ \ G^S_{b_1,b_2a}\equiv G^S_{b_1,(b_2a)}
\eqno(2.10)$$
and $E^{\mu_1\mu_2\mu_3}$ is the equation of motion of the three form gauge field which was given in references [6,2,3]. 
\par
The dual graviton equation was found by reading off the coefficient of the parameter 
$\Lambda^{c_1c_2c_3}$ in equation (2.8) and it was found to be 
$$
\hat E^{a_1\ldots a_8 }{}_{,b}\equiv    e_{\mu_1}{}^{a_1}\ldots e_{\mu_8}{}^{a_8} e_b{}^\tau \partial_{[\nu |} \{(\det e)^{{1\over 2}} G^{[\nu , \mu_1\ldots \mu_8]}{}_{,|\tau]}       \}
+\ldots 
=0
\eqno(2.11)$$
where $+\ldots $ mean terms that are constructed from the Cartan forms and are of the generic  form   $fG_{b, \bullet} $ where $f$ is  any  function of the fields of the non-linear realisation  and $G_{b, \bullet }$ is a Cartan form with $\bullet$ being any $E_{11}$ index. We refer to such terms as $l_1$ terms. The precise indices on this expression are not shown but they are to be arranged so that they are  those of the dual gravity equation of motion. 
\par
 The reason for this ambiguity  is that one can add to the six from equation of motion of equation (2.7) terms of the form $f G^{c_1c_2}{}_{,\bullet}$ where the $c_1c_2 $ indices correspond to a level one derivative with respect to the coordinate $x_{c_1c_2}$. Using equation (1.6) we find that this varies into the expression 
 $6f\Lambda^{c_1c_2 b} G_{b,\bullet }$. Looking at the variation of the six form equation (2.8) we see that this corresponds to  the ambiguity in the dual gravity equation (2.11). This ambiguity reflects  the fact that we compute the equation of motion to only lowest level in the spacetime derivatives and in order to achieve this we must include terms in the object being varied that contain terms that have derivatives with respect to the level one coordinates. 
\par
An attempt to resolve this  ambiguity was  made in reference [12] by demanding diffeomorphism invariance. This assumed that the dual graviton field transformed under a general coordinate transformation as a standard tensor as indicated by its indices. However,  this  is not the case. The dual graviton is not just any  matter field but it   describes gravity in its own way which is different to that due to the usual graviton. As a result it is to be expected that it does not behave like any other matter field. The derivation of equation (2.11) is correct if, as was stated in the paper,  one takes it to be subject to the ambiguity discussed above. However, the attempt to resolve the ambiguity by using standard diffeomorphism invariance was not correct and as a result  the  last term added to equation (2.11) in reference [12] was not correct. 
\par
Our task in this paper is to resolve the ambiguity and so derive the correct equation of motion for the dual graviton in eleven dimensions. The dual graviton equation must belong to the same representation of GL(11) as the dual graviton field. As $h_{[a_1\ldots a_8, b]}=0$ the dual graviton equation   must satisfy the condition 
$$
E_{[a_1\ldots a_8 }{}_{,b]}=0
\eqno(2.12)$$
Equation (2.11) of reference [12] does not satisfy this condition but we will, in this paper,  use the ambiguity mentioned above to make it symmetric. 
A priori, it is far from clear that this will work but we will find that it does. We will refer to quantities that we add to the dual graviton equation  that satisfy  the condition of equation (2.12) as being symmetric. 
\par
The first step is to rewrite equation (2.11) such that it involves the dual graviton Cartan form with tangent space indices, it becomes 
$$
\hat E^{a_1..a_8}{}_{b} = (\det e)^{1 \over 2} e_{[c|}{}^{\nu} \partial_{\nu} G^{[c,a_1 \ldots a_8]}{}_{,|b]} - 8 G_{[c|,e}{}^{a_1} G^{[c,e a_2 \ldots a_8]}{}_{,|b]} + G_{[c,b]}{}^e G^{[c,a_1 \ldots a_8]}{}_{,e} $$
$$
-G_{[c,|e}{}^{c} G^{[e,a_1 \ldots a_8]}{}_{,|b]} + {1 \over 2} G_{[c|,e}{}^{e} G^{[c,a_1 \ldots a_8]}{}_{,|b]} \equiv \hat E^{(1)}{}^{a_1\ldots a_8 }{}_{,b} + \hat E^{(2)}{}^{a_1\ldots a_8 }{}_{,b}
\eqno(2.13)$$
We have written the dual graviton equation as a sum of two parts, the first of which ($\hat E^{(1)}{}^{a_1\ldots a_8 }{}_{,b} $) is of the generic  form $\partial G_{1,8, 1}$ and the second of which ($\hat E^{(2)}{}^{a_1\ldots a_8 }{}_{,b} $) is 
of the generic form $G_{1,1,1}G_{1,8,1}$ where $G_{1,8, 1}$ and $G_{1,1,1}$ denote the dual gravity and gravity Cartan forms respectively. 
{\bf As above we will  adopted the convention in this paper that the $a_1\ldots a_8$ indices contained  in any equation are always completely antisymmetrised. }
\par
We begin by processing the $G_{1,1,1}G_{1,8,1}$  terms which can be written as 
$$
\hat E^{(2)}{}^{a_1\ldots a_8 }{}_{,b}=
$$
$$ 
\left( - 4 G_{c,e}{}^{a_1} G^{[c,ea_2 \ldots a_8]}{}_{,b} + {1 \over 2} G_{c,b}{}^e G^{[c,a_1 \ldots a_8]}{}_{,e} 
 - {1 \over 2} G_{e,c}{}^{e} G^{[c,a_1 \ldots a_8]}{}_{,b} + {1 \over 4} G_{c,e}{}^{e} G^{[c,a_1 \ldots a_8]}{}_{,b}\right) $$
$$
+ \left(4 G_{b,e}{}^{a_1} G^{[c,ea_2 \ldots a_8]}{}_{,c} - {1 \over 2} G_{b,c}{}^e G^{[c,a_1 \ldots a_8]}{}_{,e} + {1 \over 2} G_{b,c}{}^{e} G^{[c,a_1 \ldots a_8]}{}_{,e} - {1 \over 4} G_{b,e}{}^{e} G^{[c,a_1 \ldots a_8]}{}_{,c}\right)
$$
$$
\equiv A1+A2+A3+A4 
$$
$$
+\left( 4 G_{b,e}{}^{a_1} G^{[c,ea_2 \ldots a_8]}{}_{,c} - {1 \over 2} G_{b,c}{}^e G^{[c,a_1 \ldots a_8]}{}_{,e} + {1 \over 2} G_{b,c}{}^{e} G^{[c,a_1 \ldots a_8]}{}_{,e} - {1 \over 4} G_{b,e}{}^{e} G^{[c,a_1 \ldots a_8]}{}_{,c} \right)
\eqno(2.14)$$
where A1, A2, A3 and A4 are the terms in the first bracket in the order in which they occur. 
\par
The terms in the second bracket in equation (2.14) contain  Cartan forms whose first index is a $b$ which is contracted with the parameter $\Lambda^{c_1c_2b}$ in the variation of the six form equation (2.8). Such terms  are $l_1$ terms  and can, as we explained above,  be removed from the dual graviton equation by adding terms to the six form equation of motion. The final term in the first bracket, the term A4, can be written as 
$$
A4=  {1 \over 4 \cdot 9} \{ G_{c,e}{}^{e}  G^{c,a_1 \ldots a_8}{}_{,b} \} + {1 \over 4 \cdot 9}  G_{c,e}{}^{e} \{  8  G^{a_1,a_2 \ldots a_8c}{}_{,b} - G_{b,}{}^{a_1 \ldots a_8,c}  \}  
$$
$$
+ {1 \over 4 \cdot 9} \bigg[ G_{c,e}{}^{e} G_{b,}{}^{a_1 \ldots a_8,c }\bigg]
\eqno(2.15)$$
The first term in this equation is obviously symmetric as $G_{c, [a_1\ldots a_8, b]}=0$ as the dual graviton field and its corresponding generator satisfy this irreducibility condition.  {\bf We reserve the use of $\{ \}$ brackets to denote quantities that are symmetric as this will make it easier to keep track of them.}  The second term in this equation is also symmetric as the expression in the bracket can also be written as  
$$
8  G^{a_1,a_2 \ldots a_8 c}{}^{,b} - G^{b,}{}^{a_1 \ldots a_8,c}  = 8  G^{a_1,a_2 \ldots a_8 c}{}^{,b} - 8 G^{b,}{}^{a_1 \ldots a_7 c , a_8}  
\eqno(2.16) $$
and taking antisymmetry in the indices $a_1, \ldots a_8$ and $b$ it obviously vanishes. We have switched the position of the $c$ index using the irreducibility condition on the dual graviton Cartan form. The final term in equation (2.15) can be removed from the dual graviton equation of motion as it is an $l_1$ term. {\bf We are placing all $l_1$ terms in square brackets so that it will be easier to keep track of them. }
\par
The third term in the first bracket, the term A3,  in equation (2.14) can also be written in a very similar way, namely 
$$
A3= - {1 \over 2 \cdot 9}  \{G_{e,c}{}^{e}  G^{c,a_1 \ldots a_8}{}_{,b} \}- {1 \over 2 \cdot 9} \{ G_{e,c}{}^{e}  8  G^{a_1,a_2 \ldots a_8c}{}_{,b} - G_{b,}{}^{a_1 \ldots a_8,c}  \}  
$$
$$
-{1 \over 2 \cdot 9}\bigg[  G_{c,e}{}^{e} G_{b,}{}^{a_1 \ldots a_8,c }\bigg]
\eqno(2.17)$$
The first two terms are symmetric and the last term is an $l_1$ term. The first and second terms in equation (2.14),  that is, the terms A1 and A2,  are not symmetric and we will return to them later.
\par
We will now analyse the $\partial G_{1,8, 1}$ terms which are contained in $\hat E^{(1)}{}^{a_1\ldots a_8 }{}_{,b}$  in equation (2.13). We can write   these terms as 
$$
\hat E^{(1)}{}^{a_1\ldots a_8 }{}_{,b}= 
{1 \over 2 \cdot 9} \{ (\det e)^{1 \over 2} e_c{}^{\mu} \partial_{\mu} G^{c,a_1 \ldots a_8}{}_{,b} \}
$$
$$
+ {1 \over 2 \cdot 9}\{ (\det e)^{1 \over 2} e^c{}^{\mu} \partial_{\mu} (8 G_{a_1,a_2 \ldots a_8c}{}_{,b} - G_{b,}{}_{a_1 \ldots a_8}{}_{,c}) \}
$$
$$
- {4 \over 9}\{ (\det e)^{1 \over 2}( e_b{}^{\mu} \partial_{\mu} G^{a_1,a_2 \ldots a_8c}{}_{,c} - e_{[b}{}^{\mu} \partial_{\mu} G^{a_1,a_2 \ldots a_8]c}{}_{,c})\}
$$
$$
- {1 \over 2 \cdot 9} (\det e)^{1 \over 2}( e_b{}^{\mu} \partial_{\mu} G_{c,}{}^{a_1 \ldots a_8,c} - e_c{}^{\mu} \partial_{\mu} G_{b,}{}^{a_1 \ldots a_8,c})
$$
$$
-  {4 \over 9 \cdot 9} (\det e)^{1 \over 2} ( e_{b}{}^{\mu} \partial_{\mu} G^{a_1,a_2 \ldots a_8c}{}_{,c} - e^{a_1}{}^{\mu} \partial_{\mu} G^{b,a_2 \ldots a_8c}{}_{,c}) $$
$$
-  {4 \cdot 7 \over 9 \cdot 9} (\det e)^{1 \over 2} e^{a_1}{}^{\mu} \partial_{\mu} G^{a_2,a_3 \ldots a_8 b c}{}_{,c}  
\eqno(2.18)$$
The first and third terms in equation (2.18)  are obviously symmetric. The second terms is also symmetric for the same  reason as outlined  in equation (2.16).   However, the remaining three terms are not symmetric. 
\par 
Let us consider the fourth term  which we can rewrite as 
$$
- {1 \over 2 \cdot 9} (\det e)^{1 \over 2}\bigg[  e_b{}^{\mu} \partial_{\mu} G_{c,}{}^{a_1 \ldots a_8,c} \bigg]- e_c{}^{\mu} \partial_{\mu} G_{b,}{}^{a_1 \ldots a_8,c})
$$
$$
= - {1 \over 2 \cdot 9} (\det e)^{1 \over 2}\bigg[( e_b{}^{\mu} \partial_{\mu} G_{c,}{}^{a_1 \ldots a_8,c}  \bigg] - (e_c{}^{\mu} \partial_{\mu} e_b{}^\nu )G_{\nu,}{}^{a_1 \ldots a_8,c} - \bigg[ e_b{}^\nu (e_c{}^{\mu}) \partial_{\mu} G_{\nu,}{}^{a_1 \ldots a_8,c} \bigg] 
\eqno(2.19)$$
 For a term to be  an $l_1$ term it must,  when multiplied by the parameter $\Lambda^{c_1c_2 b} $ that arises in the variation of the six form equation (2.8),  contain a factor  of the form $\Lambda^{c_1c_2 b} G_{b, \bullet}$.  The first term in the first  line is an $l_1$ term and so can be removed from the dual gravity equation. However, the second term in the first   line is not of this form as there is a derivative  in between the parameter $\Lambda^{c_1c_2 b}$ and the Cartan form $ G_{b, \bullet}$.  As a result we have rewritten the expression in the second line. Here the  first term is an  $l_1$ terms and  so is the third term. If  we multiply this term by the parameter 
 $\Lambda^{c_1c_2 b}$ we find the factor $ e_{\rho_1} {}^{c_1}   e_{\rho_1} {}^{c_2} \Lambda^{\rho_1\rho_2 \nu} $ and using the fact that the parameter with upper world indices is a constant we can take it past the derivative to find that that the  term  is indeed an $l_1$ term. Left over from the term of equation (2.19) is the second term which can be written as 
 $$
M10\equiv -{1\over 2.9} G_{c,b}{}^e G_{e , }{}^{a_1\ldots a_8}  
\eqno(2.20)$$
This term  will be needed  in the calculation  later on. 
 \par
 The fifth term   in equation (2.18) can be treated in a similar way and we find that   
 $$
- {4 \over 9\cdot 9}(\det e)^{1 \over 2}\bigg( \bigg[  e_b{}^{\mu} \partial_{\mu} G^{a_1,}{}^{a_2 \ldots a_8c,}{}_{c} \bigg]- \bigg[ e_b{}^{\mu}e^{a_1}{}^\nu \partial_{\nu} G_{\mu,}{}^{a_2 \ldots a_8 c,}{}_{c})\bigg] \Bigg)- {4 \over 9\cdot 9}G^{a_1}{}_{ ,b e} G^{e, a_2\ldots a_8 c,}{}_{c}
$$
$$
\equiv  - {4 \over 9\cdot 9}(\det e)^{1 \over 2}\bigg( \bigg[ e_b{}^{\mu} \partial_{\mu} G^{a_1,}{}^{a_2 \ldots a_8c,}{}_{c} \bigg]- \bigg[ e_b{}^{\mu}e^{a_1}{}^\nu \partial_{\nu} G_{\mu,}{}^{a_2 \ldots a_8 c,}{}_{c})\bigg] \Bigg)+M11
 \eqno(2.21)$$
 In this equation we find two $l_1$ terms and one term, denoted as M11,  which will be needed later. 
\par
The last term in equation (2.18) can not be analysed in this way and we now use the fact that it can  be further evaluated  using the Maurer-Cartan equations of $E_{11}$  for the dual graviton Cartan form which we now derive . 
The Cartan forms of $E_{11}$ are contained in the expression ${\cal V} =g_E^{-1} d g_E$ where $g_E$ is the group element of $E_{11}$. It obviously obeys the  usual Maurer-Cartan equation $d{\cal V} +{\cal V} \wedge {\cal V} =0$. The precise expression for ${\cal V}$ in terms of the Cartan forms has been  discussed in, for example, in references [2] and [12]. Using the $E_{11}$ algebra one can show that the last term in equation (2.18) is given by 
$$
-{4.7\over 9.9}(\det e)^{1 \over 2} e_{[a_1}{}^{\mu}\partial_\mu G_{a_2 \ldots a_8]bc,}{}^c = - {4.7\over 9.9} \bigg( {1 \over 2} G_{a_1 ,e}{}^e G_{a_2, a_3 \ldots a_8 bc,}{}^{c}- G_{a_1,a_2}{}^e G_{e,a_3 \ldots a_8 bc,}{}^{c} 
$$
$$- {8 \cdot 8 \over 9} G_{a_1,[a_3 |}{}^e G_{a_2,e|a_4 \ldots a_8 bc],}{}^c 
+ {8 \over 9} G_{a_1,}{}^{c}{}^e G_{a_2,e[a_3 \ldots a_8b,c]} - {7 \cdot 8 \over 9} G_{a_1,[a_3|}{}^e G_{a_2,e|a_4 \ldots a_8b}{}^{c}{}_{,c]} 
$$
$$
- {8 \over 9} G_{a_1,}{}^{ce} G_{a_2,[a_3 \ldots a_8 bc],e}  + {8 \over 9} G_{a_1,[a_3|}{}^e G_{a_2|,a_4 \ldots a_8bc]}{}^{c}{}_{,e} 
$$
$$
- 2 G_{a_1,[a_3 a_4 a_5} G_{|a_2|,a_6 a_7 a_8 bc]}{}^c + 2 G_{a_1,[a_3 a_4|}{}^c G_{a_2|,a_5 \ldots a_8 bc]} \bigg)
$$
$$\equiv M1+M2+\ldots M8+M9
\eqno(2.22 )$$
where $M1,\ldots M9$ denote  the expressions in the order in which they occur. The reader may like to  analyse the third and fourth terms of equation (2.18)  using the Maurer-Cartan equations to recover the same results as stated above. 
\par
Our next  task is to evaluated the terms in equation (2.22).  Let us first consider the first  term in equation (2.22), that is, the terms $M1$. We may rewrite this term as  
$$
M1=- {2.7\over 9.9} G_{a_1 ,e}{}^e G_{a_2, a_3 \ldots a_8 bc,}{}^{c}
$$
$$
=- {2.7\over 9.9}\{ G_{a_1 ,e}{}^e G_{a_2, a_3 \ldots a_8 bc,}{}^{c} 
 - {1 \over 2} G_{b,e}{}^e G_{a_1,a_2 \ldots a_8c,}{}^c + {1 \over 2} G_{a_1,e}{}^e G_{b,a_2 \ldots a_8c,}{}^c\}
 $$
 $$+ {2.7\over 9.9}\big[{1 \over 2} G_{b,e}{}^e G_{a_1,a_2 \ldots a_8c,}{}^c - {1 \over 2} G_{a_1,e}{}^e G_{b ,a_2 \ldots a_8c,}{}^c\big]
\eqno(2.23)$$
As the curly brackets indicate the first term is symmetric while the terms in the last line are $l_1$ terms and can be removed. To see that the first  term is symmetric we note that we can write it as 
$$
 \{G_{a_1 ,e}{}^e G_{a_2, a_3 \ldots a_8 bc,}{}^{c} 
 - {1 \over 2} G_{b,e}{}^e G_{a_1,a_2 \ldots a_8c,}{}^c + {1 \over 2} G_{a_1,e}{}^e G_{b,a_2 \ldots a_8c,}{}^c\}
 $$
 $$
 = {9\over 2}\{ G_{a_1 ,e}{}^e G_{a_2, a_3 \ldots a_8 bc,}{}^{c} - G_{[a_1 |,e}{}^e G_{|a_2, a_3 \ldots a_8 b ]c,}{}^{c} \}
  \eqno(2.24)$$
\par
The second term in equation (2.22), that is, the term M2 can be combined with the term M11 of equation (2.21) to give the result 
$$
M2 + M11= 
 {4 \over 9 \cdot 9} \{7 G_{[a_1,a_2}{}^e G_{|e|,a_3 \ldots a_8]bc,}{}^c - G_{[a_1,|b}{}^e G_{e|,a_2 \ldots a_8]c,}{}^c - 8 G_{b,[a_1}{}^e G_{|e|,a_2 \ldots a_8]c,}{}^c\}
$$
$$
+{4.8 \over 9 \cdot 9} \bigg[ G_{b,[a_1}{}^e G_{|e|,a_2 \ldots a_8]c,}{}^c\bigg]
 \eqno(2.25) $$
The first term is symmetric as can be verified along the lines used in equation (2.24) and the last term is an $l_1$ term that can be removed. 
\par
Let us now consider the sixth and seventh  terms in equation (2.22), that is, the terms $M6+M7$,  which can be written as 
$$
M6+M7= -{4.7\over 9.9} \{- {8 \over 9} G_{a_1,}{}^{ce} G_{a_2,[a_3 \ldots a_8 bc],e}  + {8 \over 9} G_{a_1,[a_3|}{}^e G_{a_2|,a_4 \ldots a_8bc]}{}^{c}{}_{,e} \}
$$
$$
= +{4.7\over 9.9}  G_{a_1,c}{}^e G_{a_2,a_3 \ldots a_8bc}{}^{c}{}_{,e} 
$$
$$= + {4 \cdot 7 \over 9 \cdot 9} \{G_{a_1,c}{}^e G_{a_2,a_3 \ldots a_8bc,e} - {1 \over 2} G_{b,c}{}^e G_{a_1,a_2 \ldots a_8c,e} + {1 \over 2} G_{a_1,c}{}^e G_{b,a_2 \ldots a_8c,e}\}
$$
$$+ {2 \cdot 7 \over 9 \cdot 9}\bigg[ G_{b,c}{}^e G_{a_1,a_2 \ldots a_8c,e} -G_{a_1,c}{}^e G_{b,a_2 \ldots a_8c,e}\bigg]
\eqno(2.26)$$
The terms in the first bracket are symmetric and the terms in the second line can be removed as they are $l_1$ terms. 
\par
The third, fourth and fifth terms in equation (2.22) can be evaluated as follows 
$$
M3+M4+M5= - {4 \cdot 7 \over 9 \cdot 9} \big(6 G_{a_1,a_2}{}^e G_{a_3,a_4 \ldots a_8bec,}{}^c - G_{a_1,b}{}^e G_{a_2,a_3 \ldots a_8ec,}{}^c  +G_{a_1,c}{}^e G_{a_2,a_3 \ldots a_8eb,}{}^c \big)
$$
$$
=- {4 \cdot 7 \over 9 \cdot 9} \{6 G_{a_1,a_2}{}^e G_{a_3,a_4 \ldots a_8bec,}{}^c - G_{a_1,b}{}^e G_{a_2,a_3 \ldots a_8ec,}{}^c 
- 8 G_{b,a_1}{}^e G_{a_2,a_3 \ldots a_8ec,}{}^c 
$$
$$
+ G_{a_1,a_2}{}^e G_{b,ea_3 \ldots a_8c,}{}^c\} 
$$
$$
 +{4 \cdot 7 \over 9 \cdot 9} \{G_{a_1,c}{}^e G_{a_2,a_3 \ldots a_8be,c} - {1 \over 2} G_{b,c}{}^e G_{a_1,a_2 \ldots a_8e,c} + {1 \over 2} G_{a_1,c}{}^e G_{b,a_2 \ldots a_8e,c}\}
 $$
 $$
  + {4 \cdot 7 \over 9 \cdot 9} \big[- 8 G_{b,a_1}{}^e G_{a_2,a_3 \ldots a_8ec,}{}^c + G_{a_1,a_2}{}^e G_{b,ea_3 \ldots a_8c,}{}^c
  + {1 \over 2} G_{b,c}{}^e G_{a_1,a_2 \ldots a_8e,c} 
  $$
  $$
  - {1 \over 2} G_{a_1,c}{}^e G_{b,a_2 \ldots a_8e,c} \big]
\eqno(2.27)$$
The terms in curly brackets are symmetric and the remaining terms are $l_1$ terms that must be removed. 
\par
Finally the eigth and ninth terms in equation (2.22) can be evaluated as 
$$
M8+M9 = 
- {1 \over 9}\{- 2 G_{a_1,ba_2}{}^c G_{a_3,a_4 \ldots a_8}{}^c + 5 G_{a_1,a_2 a_3}{}^c G_{a_4, a_5 \ldots a_8bc} 
$$
$$- 8 G_{b,a_1 a_2}{}^{c} G_{a_3,a_4 \ldots a_8c} - G_{a_1,a_2 a_3}{}^{c} G_{b,a_4 \ldots a_8c}\} 
- {1 \over 9}\big[ 8 G_{b,a_1 a_2}{}^{c} G_{a_3,a_4 \ldots a_8c} + G_{a_1,a_2 a_3}{}^{c} G_{b,a_4 \ldots a_8c}\big]
\eqno(2.28)$$
Where the first term is symmetric and the last term is an $l_1$ that can be removed. We note that the term 
 \par
The only terms we have not processed so far are the terms A1 and A2 of equation (2.14) and the term M10 of equation (2.20). We find that the $A2 +M10$ can be written as 
$$
A2+M10= G^{[c,}{}_b{}^{e]} G_{[c, a_1, e a_2\ldots a_8],e} - {4 \over 9}  \{G^{c,}{}_{b}{}^{e} G_{a_1,e a_2 \ldots a_8,c}
+G^{c,}{}_{a_1}{}^{e} G_{b,ea_2 \ldots a_8,c}\} 
$$
$$+{4 \over 9} \bigg[ G^{c,}{}_{a_1}{}^{e} G_{b,ea_2 \ldots a_8,c}\bigg]
\eqno(2.29)$$
In this equation,  the first term we will be needed later, the middle term is a symmetric term and the last term is an $l_1$ term that we will remove. 
\par
The first term, A1, of equation (2.14) can be written as 
$$
A1 = - 4 G_{[c,e]a_1} G^{[c,e a_2 \ldots a_8]}{}_{,b} 
=  -4(G_{[c,e]a_1} + G_{[c|,a_1|e]}) G^{[c,e a_2 \ldots a_8]}{}_{,b}  + 4 G_{[c|,a_1|e]} G^{[c,e a_2 \ldots a_8]}{}_{,b}  $$
$$
\equiv A1.1 + A1.2 
\eqno(2.30)$$
We can  reformulate the first term in equation (2.30) to be given by 
$$
A1.1=  4 (E^{a_1,}{}_{ce} -G^{a_1,}{}_{[ce]}) G^{[c,e a_2 \ldots a_8]}{}_{,b} + \{ \varepsilon_{ce}{}^{f_1 \ldots f_9} G_{[f_1,f_2 \ldots f_9]}  G^{[c,e a_2 \ldots a_8]}{}_{,b} \}
\eqno(2.31)$$
where  the gravity-dual gravity relation is given in equation (2.9). The last term is in fact symmetric. 
\par
After some work the second term (A.1.2) in equation (2.30) can be rewritten as 
$$
A1.2=  -G^{[c,}{}_b{}^{e]} G_{[c,a_1 \ldots a_8],e} + {4 \over 9} \{\big(G_{c,a_1}{}^{e} G^{c,}{}_{ea_2 \ldots a_8,b} - G_{e,a_1}{}^{c} G_{c,}{}^e{}_{a_2 \ldots a_8,b}\big)
$$
$$
 +{1 \over 8 }( G_{c,b}{}^e G^{c,}{}_{a_1 \ldots a_8,e} - G_{e,b}{}^c G_{c,a_1 \ldots a_8,}{}^{e})\}
 $$
 $$
 +  \{{7 \cdot 4 \over 9} G^{[c,}{}_{a_1}{}^{e]} G_{a_2,cea_3 \ldots a_8,b} - {8 \over 9} G_{[c|,b}{}^{|e]} G_{a_1,ca_2 \ldots a_8}{}_{,e} \} 
$$
$$
+{7 \cdot 4 \over 9\cdot 9}\bigg[  G^{[c,}{}_{a_1}{}^{e]} G_{b,cea_2 \ldots a_7, a_8} \bigg]- {8 \over 9\cdot 9 } \bigg[ G_{[c |, a_1 |e]} G_{b,ca_2 \ldots a_8}{}_{,e} \bigg]
\eqno(2.32)$$
\par
Examining equation (2.32) we see that the first term cancels the first  term in equation (2.29).  The two  terms in curly brackets are symmetric and the terms in the final two bracket are $l_1$ terms which can be removed. It may not be immediately apparent to the reader that the second term in curly brackets really is symmetric. This becomes obvious if one uses the irreducibility of the dual gravity Cartan form and in particular the identity 
$$
 G_{a_1,ca_2 \ldots a_8}{}_{,e} -  G_{a_1,ea_2 \ldots a_8}{}_{,c}=  7G_{a_1,cea_2 \ldots a_7,a_8}
  \eqno(2.33)$$
\par 
In addition to the above terms there are  terms which we can add to the dual graviton equation of motion which are symmetric but at the same time are $l_1$ terms. Such terms must contain a $b$ index as the first index on one of the two Cartan forms and an $a_1$ as the first index on the other Cartan. To saturate the remaining eleven indices we also need two summed over indices. The possible terms are 
$$
c_1 (G_{b, e}{}^e G_{a_1, a_2\ldots a_8 c,}{}^{c} + G_{a_1, e}{}^e G_{b,  a_2\ldots a_8 c,}{}^{c} )
 $$
 $$
 c_2 (G_{b,}{}^{ e}{}^c G_{a_1, a_2\ldots a_8 (e,}{}_{c)} + G_{a_1, }{}^{ e}{}^c  G_{b,  a_2\ldots a_8  (e,}{}_{c)} )
 $$
 $$
 c_3 (G_{b,}{}^{ e}{}^c G_{a_1, a_2\ldots a_8 [e,}{}_{c]} + G_{a_1, }{}^{ e}{}^c  G_{b,  a_2\ldots a_8  [e,}{}_{c]} )
 $$
 $$
  c_4 (G_{b,a_2}{}^c G_{a_1, a_3\ldots a_8 ce,}{}^{e} + G_{a_1,a_2}{}^c G_{b, a_3\ldots a_8 ce,}{}^{e} )
   \eqno(2.34)$$
  where $c_1 , \ldots ,c_4$ are constants. 
 %%%%%%%%%%%%
\medskip
{\bf 3. The dual gravity equation of motion}
\medskip
To find the dual graviton equation we just need to collect up all the symmetric parts given in the curly brackets in the previous section. To make the expression self contained we will write the so far surpressed  antisymmetry on the $a$ indices and drop the curly brackets. The dual graviton equation is given by 
$$
E_{a_1 \ldots a_8}{}_{,b} \equiv {1 \over 9 \cdot 2} (\det e)^{1 \over 2} e^c{}^{\mu} \partial_{\mu} G_{c,a_1 \ldots a_8}{}_{,b} + {1 \over 2 \cdot 9} (\det e)^{1 \over 2} e^c{}^{\mu} \partial_{\mu} (8 G_{[a_1,a_2 \ldots a_8]c}{}_{,b} - G_{b,}{}_{a_1 \ldots a_8}{}_{,c})$$
$$
- {4 \over 9 \cdot 9}  (\det e)^{1 \over 2} ( 8  e_b{}^{\mu} \partial_{\mu} G_{[a_1,a_2\ldots a_8]c,}{}^{c} + e_{[a_1|}{}^{\mu} \partial_{\mu} G_{b|,}{}_{a_2 \ldots a_8]c,}{}^{,c} - 7  e_{[a_1}{}^{\mu} \partial_{|\mu|} G_{a_2,a_3 \ldots a_8]bc,}{}^{c}) $$
$$
- {2 \cdot 7 \over 9 \cdot 9} (G_{[a_1,|e|}{}^e G_{a_2,a_3 \ldots a_8]bc,}{}^c - {1 \over 2} G_{b,e}{}^e G_{[a_1,a_2 \ldots a_8]c,}{}^c + {1 \over 2} G_{[a_1,|e}{}^e G_{b|,a_2 \ldots a_8]c,}{}^c) $$
$$
+ {1 \over 9 \cdot 4} G^{c,e}{}_e G_{c,a_1 \ldots a_8,b} + {1 \over 9 \cdot 4} G^{c,e}{}_e (8G_{[a_1,a_2 \ldots a_8] c,b} - G_{b,a_1 \ldots a_8,c})$$
$$
- {1 \over 2 \cdot 9} G^{e,c}{}_e (G_{c,a_1 \ldots a_8,b} + 8 G_{[a_1,a_2 \ldots a_8]c,b} - G_{b,a_1 \ldots a_8,c})$$
$$
 +{4 \cdot 7 \over 9 \cdot 9} \{(G_{a_1,}{}^{c}{}^e +G_{a_1,}{}^{e}{}^c)G_{a_2,a_3 \ldots a_8be,c} - {1 \over 2} (G_{b,}{}^{c}{}^e+G_{b,}{}^{e}{}^c) G_{a_1,a_2 \ldots a_8e,c} 
 $$
 $$
 + {1 \over 2} (G_{a_1,}{}^{c}{}^e+G_{a_1,}{}^{e}{}^c) G_{b,a_2 \ldots a_8e,c}\}
 $$
 $$
- {1 \over 9}(- 2 G_{[a_1,|b|a_2}{}^c G_{a_3,a_4 \ldots a_8]}{}^c + 5 G_{[a_1,a_2 a_3}{}^c G_{a_4, a_5 \ldots a_8]bc} - 8 G_{b,[a_1 a_2}{}^{c} G_{a_3,a_4 \ldots a_8]c} $$
$$
- G_{[a_1,a_2 a_3}{}^{c} G_{|b|,a_4 \ldots a_8]c}) $$
$$
+ {4 \over 9 \cdot 9} (7 G_{[a_1,a_2}{}^e G_{|e|,a_3 \ldots a_8]bc,}{}^c - G_{[a_1,|b}{}^e G_{e|,a_2 \ldots a_8]c,}{}^c - 8 G_{b,[a_1}{}^e G_{|e|,a_2 \ldots a_8]c,}{}^c)$$
$$
+ \varepsilon^{c_1 c_2 e_1 \ldots e_9} G_{e_1,e_2 \ldots e_9,[a_1 |} G_{[c_1,c_2 |a_2 \ldots a_8]],b} $$
$$
- {4 \cdot 7 \over 9 \cdot 9} (6 G_{[a_1,a_2}{}^e G_{a_3,a_4 \ldots a_8]bec,}{}^c - G_{[a_1,|b|}{}^e G_{a_2,a_3 \ldots a_8]ec,}{}^c - 8 G_{b,[a_1}{}^e G_{a_2,a_3 \ldots a_8]ec,}{}^c $$
$$
+ G_{[a_1,a_2}{}^e G_{|b,e|a_3 \ldots a_8]c,}{}^c) $$
$$
- {4 \over 9} G_{c,[a_1}{}^e G_{|b,e|a_2 \ldots a_8],}{}^c 
+ {4 \over 9} (G_{c,[a_1}{}^{e} G^{c,}{}_{|e|a_2 \ldots a_8],b} - G_{e,[a_1}{}^{c} G_{|c|,}{}^e{}_{a_2 \ldots a_8],b}) $$
$$
+{1 \over 9 \cdot 2}( G_{c,b}{}^e G^{c,}{}_{[a_1 \ldots a_8],e} - G_{e,b}{}^c G_{c,a_1 \ldots a_8,}{}^{e})
+ {7 \cdot 4 \over 9} G^{[c,}{}_{[a_1}{}^{e]} G_{a_2,|ce|a_3 \ldots a_8],b} 
$$
$$
+ {4 \over 9} G_{c,b}{}^e G_{[a_1,a_2 \ldots a_8]}{}^{c}{}_{,e}=0
 \eqno(3.1)$$
 While it is not immediately apparent, this dual gravity equation does indeed  give the correct equation for the dual graviton field at the linearised level. This  equation  was already derived from the $E_{11}$ viewpoint in reference [16] by varying the six form equation of motion. 
\par
Under a Lorentz  transformations the Cartan from transform as 
$$
\delta \bar G_{a,b_1\ldots b_8, c}= \Lambda _a{}^e \bar G_{e,b_1\ldots b_8,c}+\Lambda _{b_1} {}^e \bar G_{a,eb_2\ldots b_8,c}+\ldots 
+\Lambda _{b_8} {}^e \bar G_{a,b_1\ldots b_7 e,c}+
\Lambda _c{}^e \bar G_{a,b_1\ldots b_8,e} , 
$$
$$ 
\delta G_{a,bc}= \Lambda _a{}^e  G_{e,bc}+\Lambda _b{}^e  G_{a,ec}+\Lambda _c{}^e  G_{a,be} + e_a{}^\mu \partial_\mu \Lambda^{cb}
\eqno(3.2)$$
It is straight forward to verify that dual graviton equation (3.1) is indeed Lorentz invariant. Carrying this out one realises that this is a very stringent check.  This calculation works without needing the terms of equation (2.34). In fact the first two of these terms are Lorentz covariant and so they are not excluded and one should consider them as added to the dual graviton equation (3.1). 
\par
As we have discussed varying the six form equation of motion we can find the dual gravity equation of motion. However, this equation is only determined up to  the presence of certain terms. In this paper we have resolved this ambiguity by demanding that the dual graviton equation have the same symmetries as the dual graviton field. The $l_1$ terms we have added, or subtracted,  to  the dual gravity equation in order to make it symmetric are contained in the terms in the square brackets given  in the previous section. To complete the calculation we have to list the changes to the six form equation (2.6)  that result  to these terms  in the dual graviton equation through  the variation of the six form equation of motion given in equation (2.8). The $l_1$  terms contain derivatives with respect to the level one derivatives and so they do not change the parts of the six form equation that contain only derivative with respect to the usual coordinates of spacetime, that is, the part we are familiar with. 
\par 
Let us give an example, the final term in equation (2.15) is such an $l_1$ term, and this occurs in the first term of the  right-hand side of the variation of the six form equation (2.8)   as
$$
+ {432}\cdot {1 \over 4 \cdot 9} \Lambda_{c_1 c_2 c_3} G_{c,e}{}^e G^{c_3,a_1 \ldots a_6 c_1 c_2,d} \eqno(3.2)$$
Thus term will be removed  in the dual graviton equation of motion by adding to six form equation equation (2.7) a term with level two derivatives whose variation under (2.6) is equal to this term with the opposite sign. The result is that we must add to the six form equation of motion (2.7) the term
$$
- 2 G_{d,e}{}^e G_{c_1 c_2,}{}^{a_1 \ldots a_6 c_1 c_2,d} . \eqno(3.3)$$
The coefficient $-2=-432\cdot {1 \over 4 \cdot 9} \cdot {1\over 6}$ where the one over six comes from the variation in equation (2.6). 
\par 
A similar procedure holds for every $l_1$ term that arose throughout section 2.The resulting $l_1$ extension of the six form equation of motion (2.7) is given by
$$
{\cal E}^{a_1 \ldots a_6} = \hat {\cal E}^{a_1 \ldots a_6} + {432 \over 6}\Big( - 4  G_{c_1 c_2,e}{}^{[a_1} G^{|[d,e| a_2 \ldots a_6 c_1 c_2]]}{}_{,d} + {1 \over 2}  G_{c_1 c_2,d}{}^e G^{[d,a_1 \ldots a_6 c_1 c_2]}{}_{,e} $$
$$
- {1 \over 2} G_{c_1 c_2,d}{}^{e} G^{[d,a_1 \ldots a_6 c_1 c_2]}{}_{,e} + {1 \over 4}  G_{c_1 c_2,e}{}^{e} G^{[d,a_1 \ldots a_6 c_1 c_2]}{}_{,d}$$
$$
+ {1 \over 2 \cdot 9} (\det e)^{1 \over 2}   e_{c_1 c_2}{}^{\Pi} (\partial_{\Pi} G_{d,}{}^{a_1 \ldots a_6 c_1 c_2,d} - e_d{}^{\nu} \partial_{\nu} G_{\Pi,}{}^{a_1 \ldots a_6 c_1 c_2,d})$$
$$
+ {4 \over 9 \cdot 9} (\det e)^{1 \over 2} e_{c_1 c_2}{}^{\Pi} (\partial_{\Pi} G^{[a_1,a_2 \ldots a_6 c_1 c_2] d}{}_{,d} - e^{[a_1}{}^{|\nu|} \partial_{\nu} G_{\Pi,}{}^{a_2 \ldots a_6 c_1 c_2] d}{}_{,d})$$
$$
+ {7 \over 9 \cdot 9}  G_{c_1 c_2,e}{}^e G^{[a_1,a_2 \ldots a_6 c_1 c_2 ]d,}{}_d - {7 \over 9 \cdot 9} G^{[a_1|}{}_{,e}{}^e  G_{c_1 c_2,}{}^{|a_2 \ldots a_6 c_1 c_2 ]d,}{}_d$$
$$
- {1 \over 4 \cdot 9} G_{d,e}{}^e  G_{c_1 c_2,}{}^{a_1 \ldots a_6 c_1 c_2,d} + {1 \over 2 \cdot 9} G_{e,d}{}^e G_{c_1 c_2,}{}^{a_1 \ldots a_6 c_1 c_2,d}$$
$$
+ {4 \cdot 7 \over 9 \cdot 9} (8 G_{c_1 c_2,}{}^{[a_1|e|} G^{a_2,}{}_{e}{}^{a_3 \ldots a_6 c_1 c_2] d}{}_{,d} - G^{[a_1,a_2|e|}  G_{c_1 c_2,e}{}^{a_3 \ldots a_6 c_1 c_2] d}{}_{,d})$$
$$
- {8 \over 9}  G_{c_1 c_2, d}{}^{[a_1 a_3} G^{a_2,a_4 a_5 a_6c_1 c_2]}{}^{d} + {1 \over 9} G^{[a_1,}{}_{d}{}^{ a_3 a_4}  G_{c_1 c_2,}{}^{a_2 a_5 a_6 c_1 c_2]}{}^{d} $$
$$
- {4 \cdot 8 \over 9 \cdot 9}  G_{c_1 c_2,}{}^{[a_1|e|} G_{e,}{}^{a_2 \ldots a_6 c_1 c_2] d}{}_{,d} - {8 \over 9 \cdot 9}  G_{[d,}{}^{[a_1}{}_{e]} G_{c_1 c_2,}{}^{|d|a_2 \ldots a_6 c_1 c_2],e} $$
$$
+ {4 \cdot 7 \over 9 \cdot 9} G_{[d,}{}^{[a_1}{}_{e]} G_{c_1 c_2,}{}^{|d e| a_2 \ldots a_6 c_1 c_2], a_8} - {4 \over 9} G_{d,}{}^{[a_1}{}_{e} G_{c_1 c_2,e}{}^{a_2 \ldots a_6 c_1 c_2],d}
 $$
 $$
+ {2 \cdot 7 \over 9 \cdot 9} \big( G_{b,}{}^{ce}+G_{b,}{}^{ec} )G_{[a_1,a_2 \ldots a_8 ]c,e} -( G_{[a_1|,}{}^{ce}+G_{[a_1,|}{}^{ec}) G_{b,|a_2 \ldots a_8 ]c,e}\big) \Big). 
 \eqno(3.4)$$
\par 
Since the six form equation of motion ${\cal E}^{a_1 \ldots a_6}$ has changed so does its  variation. With the above changes its variation is given by   
$$
\delta {\cal E}^{a_1\ldots a_6}= 432\,\Lambda_{c_1c_2c_3}\,E^{a_1...a_6c_1c_2,\,c_3} + 4 \cdot 432 \Lambda_{c_1 c_2 c_3} ( E^{[a_1|,}{}_{de}  - G^{[a_1|,}{}_{[de]}  ) G^{d,e| a_2 \ldots a_6 c_1 c_2],c_3}
$$
$$
+{8\over 7}  \Lambda ^{[a_1a_2a_3} E^{a_4a_5a_6]} +{2\over 105} G_{[e_5 ,  c_1c_2c_3]} \epsilon^{a_1\ldots a_6e_1\ldots e_5} E_{e_1\ldots e_4}\Lambda^{c_1c_2c_3}
$$
$$
-{3\over 35} G_{[ e_5, e_6 c_1c_2]}\Lambda^{c_1c_2[a_1}
\epsilon^{a_2\ldots a_6 ] e_1\ldots e_6} E_{e_1\ldots e_4}
+ {1\over 420}\,\epsilon_{c_1}{}^{a_1...a_6b_1...b_4}\,\omega_{c_2,\,b_1b_2}\,E_{c_3,\,b_3b_4}\,\Lambda^{c_1c_2c_3}
\eqno(3.5)$$
\par 
This result has the same form as  in equation (2.8) but with the hats removed and an extra term involving the gravity-dual gravity relation of equation (2.9). This  extra term, which vanishes,  arises due to first part of A1.1 in equation (2.31).
\par 
Equation (3.1) considerably simplifies if we present it in terms of world indices, the result is 
$$
E_{\mu_1 \ldots \mu_8,\tau} \equiv  g^{\nu \kappa} \partial_{[\nu|} F_{[\kappa,\mu_1 \ldots \mu_8]}{}_{,|\tau]} - {1 \over 9}  g^{\nu \kappa} \hat{G}_{\tau,\rho}{}^{\rho} \hat{G}_{[\mu_1,\mu_2 \ldots \mu_8] \nu,}{}_{\kappa} - {1 \over 9}  g^{\nu \kappa}  \hat{G}_{[\mu_1|,\rho}{}^{\rho} \hat{G}_{\tau,}{}_{|\mu_2 \ldots \mu_8] \nu}{}_{,\kappa} $$
$$
+  {1 \over 2} g^{\nu \kappa} \hat{G}_{\nu,\rho}{}^{\rho}   \hat{G}_{[\kappa,\mu_1 \ldots \mu_8]}{}_{,\tau}   - {1 \over 2 \cdot 9} g^{\nu \kappa} \hat{G}_{\nu,\rho}{}^{\rho} \hat{G}_{\tau,}{}_{\mu_1 \ldots \mu_8}{}_{,\kappa}  -  \hat{G}_{\nu,}{}^{(\kappa}{}^{\nu)} \hat{G}_{[\kappa,\mu_1 \ldots \mu_8],\tau}   $$
$$
+ {1 \over  9} \hat{G}_{\nu,}{}^{(\kappa}{}^{\nu)} \hat{G}_{\tau,\mu_1 \ldots \mu_8,\kappa} + {4 \over 9} \hat{G}_{\tau,}{}^{(\nu}{}^{ \kappa)} \hat{G}_{[\mu_1,\mu_2 \ldots \mu_8] \nu,}{}_{\kappa} + {4 \over 9} \hat{G}_{[\mu_1|,}{}^{(\nu }{}^{\kappa)} \hat{G}_{\tau,}{}_{|\mu_2 \ldots \mu_8]\nu}{}_{,\kappa} $$
$$
+ (\det e)^{-1} \varepsilon^{\kappa_1 \kappa_2 \nu_1 \ldots \nu_9} \hat{G}_{\nu_1,\nu_2 \ldots \nu_9,[\mu_1|} \hat{G}_{[\kappa_1,\kappa_2| \mu_2 \ldots \mu_8]] ,\tau}  + g^{\nu \kappa} \hat{G}_{\tau, [\mu_1 \mu_2|}{}_{\nu} \hat{G}_{|\mu_3,\mu_4 \ldots \mu_8] \kappa} $$
$$
+ {1 \over 9} g^{\nu \kappa} (\hat{G}_{\nu,[\mu_1 \mu_2 \mu_3|} \hat{G}_{\tau,|\mu_4 \ldots \mu_8] \kappa} - \hat{G}_{\nu,[\mu_1 \mu_2| \kappa} G_{\tau,|\mu_3 \ldots \mu_8]} - \hat{G}_{\tau,[\mu_1 \mu_2 \mu_3|} \hat{G}_{\nu,|\mu_4 \ldots \mu_8] \kappa} $$
$$
+ \hat{G}_{\tau,[\mu_1 \mu_2 | \kappa} \hat{G}_{\nu,|\mu_3 \ldots \mu_8]} ) =0
\eqno(3.6) $$
where we have defined
$$
\hat{G}_{\tau} {}_{,\mu}{}^{\nu} =  (\partial_\tau  e_\rho{}^b) e_b{}^{\nu} , \  \hat{G}_{\tau,\mu_1 \mu_2 \mu_3} = \partial_{\tau} A_{\mu_1\mu_2\mu_3},\ \ 
$$
$$
\hat{G}_{\tau , \mu_1\ldots \mu_6 }= (\partial_{\tau} A_{\mu_1\ldots \mu_6}-A_{[\mu_1\mu_2\mu_3|} \partial_{\tau}A_{|\mu_4\mu_5 \mu_6]})
$$
$$
F_{\tau,}{}_{\mu_1\ldots \mu_8,\nu} =   (\partial_\tau h_{\mu_1\ldots \mu_8,\nu}
-A_{[\mu_1\mu_2 \mu_3|}\partial_{\tau} A_{|\mu_4\mu_5\mu_6} A_{\mu_7\mu_8] \nu} + 2 \partial_\tau A_{[\mu_1\ldots \mu_6} A_{\mu_7\mu_8] \nu} $$
$$
+2\partial_\tau A_{[\mu_1\ldots \mu_5 \nu} A_{\mu_6\mu_7  \mu_8]}) \eqno(3.7)
$$
In these definitions we have removed the $(\det e)^{{1\over 2}}$ factors from  the Cartan forms of equation (2.1) and given them world indices.
\par
$E_{11}$ contains the Kac-Moody algebra $A_8^{+++}$ which describes just gravity in eleven dimensions. As such to obtain the dual gravity equation contained in this theory one just has to set to zero the three form and six form gauge fields in equation (3.1). It would be interesting to find the diffeomorphism and dual gravity gauge transformations that leave equation (3.1) invariant. From this one could  understand the geometry  that describes a dually symmetric theory of gravity. It would be interesting to carry out the $I_c(E_{11})$ variation of the dual gravity equation (3.1) to find the non-linear level four equation of the non-linear realisation. This would also resolve if the terms of equation (2.34) are present or not.

%%%%%%%%%%%%%%%%%%%%%%%%%%%%%%%%%%%%%%%%%%%%%%%%%%%%%%%%%%%%%%%%%%%%%%%

\medskip
{\bf {Acknowledgments}}
\medskip
We wish to thank Paul Cook  for discussions. Peter West wishes to thank the SFTC for support from Consolidated grants number ST/J002798/1 and ST/P000258/1, while Keith Glennon would like to thank Kings College  for support during his PhD studies.

%%%%%%%%%%%%%%%%%%%%%%%%%%%%%%%%%%%%%%%%%%%%%%%%%%%%%%%%%%%%%%%%%%%%%%%
\medskip
{\bf {References }}
\medskip
\item{[1]} C. Montenon and D. Olive, {\it Magnetic monopoles as gauge particles}, Phys Lett {\bf 72B} (1977) 117. 
\item {[2]} A. Tumanov and P. West, {\it E11 must be a symmetry of strings and branes},  arXiv:1512.01644.
\item{[3]} A. Tumanov and P. West, {\it E11 in 11D}, Phys.Lett. B758 (2016) 278, arXiv:1601.03974.
\item{[4]} E. Cremmer and B. Julia, {\it The $N=8$ supergravity theory. I. The Lagrangian}, Phys.\ Lett.\ {\bf 80B} (1978) 48, B.\ Julia, {\it ÊGroup Disintegrations}, p.\ 331 and E. Cremmer, Ê{\it Supergravities In 5 Dimensions}, in {\it Superspace Supergravity}, eds.\ S.W.\ Hawking Êand M.\ Ro\v{c}ek, Cambridge University Press (1981);  E. Cremmer, {\it Dimensional Reduction In Field Theory And Hidden Symmetries In Extended Supergravity}, ÊPublished in Trieste Supergravity School 1981, 313.
\item{[5]}  J, Schwarz and P. West, {\it ÊSymmetries and Transformation of Chiral $N=2$ $D=10$ Supergravity}, Phys. Lett. {\bf 126B} (1983) 301.
\item{[6]} P. West, {\it $E_{11}$ and M Theory}, Class. Quant. Grav.  {\bf 18}, (2001) 4443, hep-th/ 0104081.
\item{[7]}F. Riccioni and P. West, {\it Dual fields and $E_{11}$},   Phys.Lett.
B645 (2007) 286-292,  hep-th/0612001; F. Riccioni,  D. Steele and P.West, {\it Duality Symmetries and $G^{+++}$ Theories},  Class.Quant.Grav.25:045012,2008,  arXiv0706.3659. 
\item{[8]} N. Boulanger,  Per Sundell  and P.West, {\it Gauge fields and infinite chains of dualities}, JHEP 1509 (2015) 192,  arXiv:1502.07909. 
\item{[9]} T. Curtright, {\it Generalised Gauge fields}, Phys. Lett. {\bf 165B} (1985) 304.
\item{[10]} C. Hull, {\it   Strongly Coupled Gravity and Duality}, Nucl.Phys. {\bf B583} (2000) 237, hep-th/0004195.
\item{[11]}  X. Bekaert, N. Boulanger and M. Henneaux, {\it Consistent deformations of dual formulations of linearized gravity: A no-go result } 
Phys.Rev. D67 (2003) 044010,  arXiv:hep-th/0210278. 
X.  Bekaert, N.  Boulanger and  S.  Cnockaert, {\it No Self-Interaction for Two-Column Massless Fields}, J.Math.Phys. 46 (2005) 012303, arXiv:hep-th/0407102. 
\item{[12]} A. Tumanov and and P. West, {\it  E11 and the non-linear dual graviton}, Phys.Lett. B779 (2018) 479-484, arXiv:1710.11031.
\item{[13]} K. Glennon and P. West, {\it Gravity, Dual Gravity and A1+++ }, arXiv:2004.03363
\item{[14]} P. West,{\it  A brief review of E theory}, Proceedings of Abdus Salam's 90th  Birthday meeting, 25-28 January 2016, NTU, Singapore, Editors L. Brink, M. Duff and K. Phua, World Scientific Publishing and IJMPA, {\bf Vol 31}, No 26 (2016) 1630043, arXiv:1609.06863. 
\item{[15]} P. West, {\it Introduction to Strings and Branes}, Cambridge University Press, 2012.
\item{[16]}  A. Tumanov and and P. West, {\it $E_{11}$,  Romans theory and higher level duality relations}, IJMPA, {\bf Vol 32}, No 26 (2017) 1750023,  arXiv:1611.03369

\end